\documentclass[onecolumn,showpacs,aps,floatfix,pra,superscriptaddress,nofootinbib]{revtex4}
\usepackage{color}
\usepackage{graphicx}
\usepackage{bm}
\usepackage{amsmath}
\usepackage{enumerate}

\begin{document}

\title{Quantum dynamics in a time-dependent cylindrical trap}
\author{S. V. Mousavi}
\email{vmousavi@qom.ac.ir}
\affiliation{Department of Physics, The University of Qom, P. O. Box 37165, Qom, Iran}

\begin{abstract}

Solutions to the Schr\"{o}dinger equation are examined for a
particle inside a cylindrical trap of a circular time-dependent
cross-section. Analytical expressions for energy and momentum
expectation values are derived with respect to the exact solutions;
and the adiabatic and sudden change of the boundary are discussed.
The density profile as a function of time in a given observation
point, resembles the diffraction-in-time pattern observed in a
suddenly released particle but with an enhanced fringe visibility.
Numerical computations are presented for both contracting and
expanding boxes.

\end{abstract}
 \maketitle


\section{Introduction}

Quantum mechanical motion of a particle which is subject to
time-dependent boundary conditions belongs to a main class of the
time-dependent dynamical systems. Finding exact solutions to the
Schr\"{o}dinger equation in these time-dependent systems is not an
easy task \cite{Mo-PR-1952, DoRi-AJP-1969, MaDe-PLA-1991,
Ma-JPA-1992, Mo-EPL-2012}.

Diffraction in time \cite{Mo-PR-1952}, quantum temporal oscillations
of matter waves released from a confining region, is just one of the
very interesting effects that have been seen in such systems . In
comparison with the motionless case, the visibility of the fringes
is enhanced for a beam of particles, incident from the left on a
moving wall \cite{CaMuCl-PRA-2008}.

Moshinsky's theoretical work \cite{Mo-PR-1952} has been studied for
particles which are suddenly released from a 1D box
\cite{Go-PRA-2002_CaMu-EP-2006_Mo-JPA-2010}, a spherical
\cite{Go-PRA-2003} and a cylindrical trap \cite{GoOk-AIP-2005}.

Exact solutions of the Schr\"{o}dinger equation for a particle in a
1D box with a moving wall have been found  \cite{MaDe-PLA-1991,
DoRi-AJP-1969}. Exact propagator of this problem was evaluated by
Luz and Cheng \cite{LuCh-JPA-1992} by using the semiclassical
approximation. In an investigation by Dembi\'nski et al.
\cite{DeMaPe-JPA-1995}, they derived exact classical and quantum
mechanical expressions for positions, momenta and their probability
distributions; and their asymptotic behavior were discussed for a
particle in a uniformly expanding potential well . Two-dimensional
version of a Fermi accelerator, consisting of a free particle in a
circular trap with a radius varying periodically in time, has been
extensively studied in classical and quantum mechanics
\cite{BaJoCh-PD-1995}.

It has been noted by Godoy and Okamura \cite{GoOk-AIP-2005} that
particles motion in a nanoscopic circuit can be regarded as a
quantum wave, moving inside a wave guide filled with a scattering
medium, and transient currents in these circuits can be produced by
rapidly removing (or adding) boundary walls to the guide. See
\cite{CaCaMu-PR-2009} for a recent review on quantum transients.

Having this in mind, we aim to examine the solutions of the
Schr\"{o}dinger equation for a particle in a circular trap with a
varying radius. Even though traps do not in general have infinite
wells, working out in details a case with analytical solution is
helpful and provides insight. Having explicit expressions for the
excitations and variables studied is worthwhile and is a good
reference for the experimentalist.

There are two extra motivations for this study. First, the sudden
removal of the boundary is an idealized case. Second, this study
provides possible applications to optical effects connected with the
moving mirrors which are encouraging. The investigation by Chen et
al. in \cite{ChMuCaRu-PRA-2009}, provides a good reference for
motivation for expansions with hard-wall traps. Here, one finds a
method to achieve efficient atom cooling by comparing linear and
square-root in time expansions of a 1D box. Although, the limit of
infinite velocity of the moving boundary is clearly corresponds to
the sudden removal of the wall, due to the non-relativistic nature
of the Schr\"{o}dinger equation, the speed of the wall has to be
very less than the speed of light, $u \ll c$.

\section{Exact solution}

Consider a particle with mass $\mu$ inside a cylindrical wave guide of
circular cross-section.
The cylinder has a radius $L(t) = a + ut$, which uniformly changes
with time, and its axis is taken along the $z$-direction.
The potential energy function is zero if $r < L(t)$ and
infinite otherwise. The 2D Schr\"{o}dinger equation for the transverse
wavefunction in the polar coordinates ${\bf{r}} = (\rho, \phi)$ is then
\begin{eqnarray} \label{eq: sch}
i \hbar \frac{\partial}{\partial t} \Psi({\bf{r}}, t) &=&
-\frac{\hbar^2}{2 \mu} \left[ \frac{1}{\rho}
\frac{\partial}{\partial \rho} \left( \rho \frac{\partial}{\partial
\rho} \right) + \frac{1}{\rho ^2} \frac{\partial^2}{\partial \phi^2} \right]
\Psi({\bf{r}}, t)~.
\end{eqnarray}
%

In the absence of boundary motion, $u=0$, the solutions of
eq.(\ref{eq: sch}) would be
\begin{eqnarray} \label{eq: statationary_sol}
u_{m n}({\bf{r}}) &=& \frac{\sqrt{2}}{a} ~ \frac{1}{| J_{m +
1}(x_{mn}) |} ~ J_m \left( x_{m n}  \frac{\rho}{a} \right)
\frac{e^{\pm im\phi}}{\sqrt{2\pi}}  ~,
\end{eqnarray}
with eigenvalues $ E_{m n} = \hbar^2 x^2_{m n}/(2\mu a^2) $, where
$x_{m n}$ are the roots of Bessel functions, $J_{m}(x_{m n}) = 0$
with $m = 0, 1, 2, ...$ and $n = 1, 2, 3, ...$. All the Bessel
functions with $m \neq 0$ have a zero at the origin, but one has not
to consider these zeros to have a non-zero wavefunction.
Thus, the {\it instantaneous} energy eigenfunctions and eigenvalues
would, respectively, be
\begin{eqnarray} \label{eq: instan-waves}
u_{m n}({\bf{r}}, t) &=& \frac{\sqrt{2}}{L(t)} ~ \frac{1}{| J_{m +
1}(x_{m n}) |} ~
J_m \left( x_{m n}  \frac{\rho}{L(t)} \right) \frac{e^{\pm im\phi}}{\sqrt{2\pi}} ~e^{ikz} ~,\\
E_{m n}(t) &=& \frac{\hbar^2}{2\mu} \left( k^2 + \frac{x^2_{m n}}{L^2(t)} \right) ~,
\end{eqnarray}
which satisfy the relation
\begin{eqnarray} \label{eq: instan-equation}
i\hbar \frac{\partial}{\partial t} \left(\exp \left[
-i\frac{1}{\hbar} \int_0^t dt^{\prime}~E_{m n}(t^{\prime}) \right]
u_{m n}({\bf{r}}, t) \right) &=& E_{m n}(t) \left(\exp \left[
-i\frac{1}{\hbar} \int_0^t dt^{\prime}~E_{m n}(t^{\prime}) \right]
u_{m n}({\bf{r}}, t) \right)~. \nonumber\\
\end{eqnarray}

Exact solutions of the Schr\"odinger equation (\ref{eq: sch}) for
the problem above have been found \cite{Yu-PLA-2004},
\begin{eqnarray} \label{eq: radialwave}
\Psi_{m n}(\rho, \phi, t) &=& C \frac{1}{L(t)} \exp \left[ \frac{i
\mu}{2\hbar} u \frac{\rho ^2}{L(t)} -i\frac{\hbar}{2 \mu} x^2_{m n}
\frac{t}{aL(t)} \right] J_{m} \left( x_{m n}  \frac{\rho}{L(t)}
\right) \frac{e^{\pm im\phi}}{\sqrt{2\pi}} ~.
\end{eqnarray}

Unknown coefficient $C$ is determined by the normalization
condition $\int_0^{L(t)} d\rho \rho \int d\phi |\Psi_{m n}({\bf{r}},
t)|^2 = 1$, which apart from a constant phase factor yields in $ C =
\sqrt{2} / |J_{m + 1}(x_{m n})| $, by using the orthogonality of the
Bessel functions \cite{Arfken-book-2005}
\begin{eqnarray} \label{eq: bessel-ortho}
\int_0^1 ds~s~ J_{m}(x_{m p}s) J_{m}(x_{m q}s)   &=& \frac{1}{2} [
J_{m + 1}(x_{m p}) ]^2 \delta_{pq}~.
\end{eqnarray}

Thus, exact  solutions read
\begin{eqnarray} \label{eq: norm_waves}
\Psi_{m n}(\rho, \phi, t) &=&
\exp \left[ i \alpha \xi(t) \left(\frac{\rho}{L(t)}
\right)^2 - i x^2_{m n} \frac{1-1/\xi(t)}{4 \alpha} \right]
u_{m n}(\rho, \phi, t) ~,
\end{eqnarray}
where we have introduced dimensionless quantities $\alpha = \mu a u /(2 \hbar)$ and $\xi(t) = L(t)/a$.
An extra phase factor $\exp[ikz-i\hbar k^2 t/(2\mu)]$ is introduced when
the motion in the $z$-direction is included.
Functions $\Psi_{m n}({\bf{r}}, t)$ vanish at $\rho = L(t)$, remain
normalized as the boundary moves and form a complete orthogonal set.
Thus, the general solution of eq. (\ref{eq: sch}) can be expanded in
terms of functions (\ref{eq: norm_waves}),
\begin{eqnarray} \label{eq: geral-sol_1}
\Psi({\bf{r}}, t) &=&
\sum_{m^{\prime}=0}^{\infty} \sum_{n^{\prime}=1}^{\infty}
c_{m^{\prime} n^{\prime}}\Psi_{m^{\prime} n^{\prime}}({\bf{r}}, t)~,
\end{eqnarray}
with time-independent coefficients $c_{m^{\prime} n^{\prime}}$ determined from the relation
\begin{eqnarray} \label{eq: coef_1}
c_{m^{\prime} n^{\prime}} = \int_0^a d\rho ~ \rho \int_0^{2\pi}
d\phi ~ \Psi^*_{m^{\prime} n^{\prime} }({\bf{r}}, 0) \Psi({\bf{r}},
0) ~.
\end{eqnarray}

General solution can also be written as a superposition of
instantaneous eigenfunctions,
\begin{eqnarray} \label{eq: geral-sol_2}
\Psi({\bf{r}}, t) &=& \sum_{m^{\prime}=0}^{\infty} \sum_{n^{\prime}=1}^{\infty}
d_{m^{\prime} n^{\prime}}(t) \exp{ \left[ -\frac{i}{\hbar} \int_0^t
dt^{\prime}~E_{m^{\prime} n^{\prime}}(t^{\prime}) \right] }
u_{m^{\prime} n^{\prime}}({\bf{r}}, t)
 \equiv \sum_{m^{\prime}=0}^{\infty}
\sum_{n^{\prime}=1}^{\infty} b_{m^{\prime} n^{\prime}}(t)
u_{m^{\prime} n^{\prime}}({\bf{r}}, t)~,
 \nonumber\\
\end{eqnarray}
now with time-dependent coefficients $b_{m^{\prime} n^{\prime}}(t)$
which are determined from the relation
\begin{eqnarray} \label{eq: coef_2}
b_{m^{\prime} n^{\prime}}(t) = \int_0^{L(t)} d\rho~\rho
\int_0^{2\pi} d\phi ~ u^*_{m^{\prime} n^{\prime}}({\bf{r}}, t)
\Psi({\bf{r}}, t) ~.
\end{eqnarray}
Using eqs. (\ref{eq: geral-sol_1}) and (\ref{eq: coef_2}) and the
orthonormality of functions $\exp{(\pm im\phi)}/\sqrt{2\pi}$, one
obtains
\begin{eqnarray} \label{eq: relation_coefs}
b_{m^{\prime} n^{\prime}}(t) &=& \frac{2}{| J_{m^{\prime} + 1}(x_{m^{\prime} n^{\prime}}) |}
\sum_{n^{\prime \prime}=1}^{\infty} c_{m^{\prime} n^{\prime \prime}}
\frac{1}{| J_{m^{\prime} + 1}(x_{m^{\prime} n^{\prime \prime}}) |}
\exp \left[ -i x^2_{m^{\prime} n^{\prime\prime}} \frac{1-1/\xi(t)}{4 \alpha} \right]
I^*_{m^{\prime} n^{\prime} n^{\prime \prime}}(t, \alpha)~,
\end{eqnarray}
where
\begin{eqnarray} \label{eq: integral}
I_{m^{\prime} n^{\prime} n^{\prime \prime}}(t, \alpha) = \int_0^1 ds
~ s~ e^{-i\alpha \xi(t) s^2} J_{m^{\prime}}(x_{m^{\prime}
n^{\prime}} s) J_{m^{\prime}}(x_{m^{\prime} n^{\prime \prime}} s)~,
\end{eqnarray}
with the property $I_{m n n^{\prime}}(t, -\alpha) = I^*_{m n
n^{\prime}}(t, \alpha)$.

If the particle is initially in an energy eigenstate, {\it i.e.,} $\Psi({\bf{r}}, 0) = u_{m n}({\bf{r}}, 0)$,
then
\begin{eqnarray} \label{eq: c_cof_eigen}
c_{m^{\prime} n^{\prime}} = \delta_{m m^{\prime}}
\frac{2}{| J_{m + 1}(x_{m n}) ~ J_{m + 1}(x_{m n^{\prime}}) |} I_{m n n^{\prime}}(0, \alpha)~,
\end{eqnarray}
which is not an unexpected result, since the angular momentum
operator $ L_z = -i\hbar~\partial / \partial \phi $ commutes with
the Hamiltonian of the system. For $\alpha \ll 1$, one has $I_{m n
n^{\prime}}(0, \alpha) \simeq \delta_{n n^{\prime}}
(J_{m+1}(x_{mn}))^2/2$, and thus from eq. (\ref{eq: geral-sol_1}),
one obtains $\Psi({\bf{r}}, t) \simeq \Psi_{mn}({\bf{r}}, t)$, which
is in agreement with the result of the adiabatic approximation.

For completeness we should obtain the relations for the energy
expectation value and the propagator of the problem. The expectation
value of the energy of the particle is obtained from
\begin{eqnarray} \label{eq: Energy_exp}
\langle H \rangle(t)  &=& \int_0^{L(t)}d\rho~\rho \int_0^{2\pi}
d\phi~\Psi^*({\bf{r}}, t)~i \hbar \frac{\partial}{\partial t}
\Psi({\bf{r}}, t)
\nonumber\\
&=& \sum_{m^{\prime} n^{\prime}} |d_{m^{\prime} n^{\prime}}(t)|^2 ~
E_{m^{\prime} n^{\prime}}(t) = \sum_{m^{\prime} n^{\prime}}
|b_{m^{\prime} n^{\prime}}(t)|^2 ~ E_{m^{\prime} n^{\prime}}(t)~.
\end{eqnarray}

Propagator of the problem can be constructed as follows:
\begin{eqnarray*}
|\Psi(t) \rangle &=& S(t, t^{\prime}) |\Psi(t^{\prime}) \rangle
=
\sum_{m n} \sum_{m^{\prime} n^{\prime}} | \Psi_{m n}(t) \rangle
\langle \Psi_{m n}(t) | S(t, t^{\prime}) | \Psi_{m^{\prime} n^{\prime}} (t^{\prime}) \rangle
\langle \Psi_{m^{\prime} n^{\prime}}(t^{\prime})| \Psi(t^{\prime}) \rangle
\nonumber \\
&=&
\sum_{m n} | \Psi_{m n}(t) \rangle \langle \Psi_{m n}(t^{\prime})| \Psi(t^{\prime}) \rangle ~,
\end{eqnarray*}
where $S(t, t^{\prime})$ is the time evolution operator and we have used
the fact that if the particle is in the sate $| \Psi_{m n} \rangle $
at $t^{\prime}$, it remains in that state as the wall moves, {\it{i.e.,}}
$S(t, t^{\prime}) | \Psi_{m n}(t^{\prime})\rangle = | \Psi_{m n}(t) \rangle $.
Now, we write this equation in the form
\begin{eqnarray} \label{eq: wave-function}
\Psi({\bf{r}}, t) = \int_0^a d\rho^{\prime} \rho^{\prime}
\int_0^{2\pi} d\phi^{\prime} K({\bf{r}}, t ; {\bf{r}}^{\prime},
t^{\prime}) \Psi({\bf{r}}^{\prime}, t^{\prime}) ~,
\end{eqnarray}
where we have introduced the propagator as
\begin{eqnarray} \label{eq: propagator}
K({{\bf{r}}}, t ; {{\bf{r}}}^{\prime}, t^{\prime}) &=& \sum_{m=0}^{\infty} \sum_{n=1}^{\infty}
\Psi_{m n}({\bf{r}}, t) \Psi^*_{m n}({{\bf{r}}}^{\prime}, t^{\prime})
\nonumber \\
&=&
\frac{2}{L(t) L(t^{\prime})} \sum_{m n} \frac{1}{[ J_{m + 1}(x_{m n}) ]^2}~
\nonumber \\
& & \times
\exp \left[ \frac{i\mu u}{2\hbar} \left( \frac{\rho^2}{L(t)} -  \frac{{\rho^{\prime}}^2}{L(t^{\prime})}  \right)
 -\frac{i \hbar}{2\mu} \frac{x^2_{m n}}{a}\left( \frac{t}{L(t)} -  \frac{t^{\prime}}{L(t^{\prime})} \right) \right]
\nonumber \\
& & \times
J_{m} \left( x_{m n}  \frac{\rho}{L(t)} \right)~J_{m} \left( x_{m n}  \frac{\rho^{\prime}}{L(t^{\prime})} \right)~
\times
\frac{e^{\pm im\phi}}{\sqrt{2\pi}}~\frac{e^{\mp im\phi^{\prime}}}{\sqrt{2\pi}}~.
\end{eqnarray}

Expectation value of an arbitrary operator $\hat{A}$ is given by
\begin{eqnarray} \label{eq: A_expect_u}
\langle \Psi | \hat{A} | \Psi \rangle(t) &=&
\sum_{m^{\prime}=0}^{\infty} \sum_{n^{\prime}=1}^{\infty} \sum_{m=0}^{\infty} \sum_{n=1}^{\infty}
c^*_{m^{\prime} n^{\prime}} c_{m n} \langle \Psi_{m^{\prime} n^{\prime}}(t) | \hat{A} | \Psi_{mn}(t) \rangle ~,
\end{eqnarray}
where
\begin{eqnarray}
\langle \Psi_{m^{\prime} n^{\prime}}(t) | \hat{A} | \Psi_{mn}(t) \rangle &=&
\int_0^{2\pi} d\phi \int_0^{L(t)} d\rho~ \rho~
\Psi^*_{m^{\prime} n^{\prime}}(\rho, \phi, t) ~ A(\rho, \phi) ~ \Psi_{m n}(\rho, \phi, t) ~,
\end{eqnarray}
are the matrix elements of the operator with respect to the states
$\Psi_{mn}({\bf{r}}, t)$. It must be noted that the expectation
values can also be found by expanding the wavefunction in terms of
the instantaneous energy eigenfunctions as we obtained in eq.
(\ref{eq: Energy_exp}). Our aim is to examine the uncertainty
relations. To this end, we first compute matrix elements of the
position, the momentum and the Hamiltonian operators.

The position and the momentum operators in the arbitrary radial
direction $ \hat{\rho}_0 = \hat{x} \cos \phi_0  + \hat{y} \sin
\phi_0$, are
\begin{eqnarray}
q_{_0} &=& x \cos \phi_0 + y \sin \phi_0~,\nonumber\\
p_{_0} &=& p_x \cos \phi_0 + p_y \sin \phi_0 = \frac{\hbar}{i}
\left[ \cos \phi_0 \left( \cos \phi \frac{\partial}{\partial \rho} -
\frac{1}{\rho} \sin \phi \frac{\partial}{\partial \phi} \right) +
\sin \phi_0 \left( \sin \phi \frac{\partial}{\partial \rho} +
\frac{1}{\rho} \cos \phi \frac{\partial}{\partial \phi} \right)
\right] ~,
\end{eqnarray}
which are canonical conjugate variables, $[q_{_0}, p_{_0}] =
i\hbar$. By straightforward algebra, one obtains
\begin{eqnarray}
\langle \Psi_{m^{\prime} n^{\prime}}(t) | q_{_0}  | \Psi_{mn}(t) \rangle &=& 0~, \label{eq: q0_expect}
\\
\langle \Psi_{m^{\prime} n^{\prime}}(t) | p_{_0}  | \Psi_{mn}(t) \rangle &=& 0 ~, \label{eq: p0_expect}
\\
\langle \Psi_{m^{\prime} n^{\prime}}(t) | q^2_{_0}  | \Psi_{mn}(t) \rangle \rangle &=&
\delta_{m, m^{\prime}} \exp \left[ i ( x^2_{m n^{\prime}} -  x^2_{m n}) \frac{1-1/\xi(t)}{4 \alpha} \right]
a^2 \xi^2(t) \frac{2 \pi A^{(3)}_{m n^{\prime} n}}{ |J_{m+1}(x_{mn^{\prime}}) ~ J_{m+1}(x_{mn}) | } ~, \label{eq: q0^2_expect}
\\
\langle \Psi_{m^{\prime} n^{\prime}}(t) | p^2_{_0}  | \Psi_{mn}(t) \rangle &=&
\delta_{m, m^{\prime}} \exp \left[ i ( x^2_{m n^{\prime}} -  x^2_{m n}) \frac{1-1/\xi(t)}{4 \alpha} \right]
\frac{\hbar^2}{a^2} \frac{2 \pi}{| J_{m+1}(x_{mn^{\prime}}) ~ J_{m+1}(x_{mn}) |}
\nonumber
\\
& \times &
\left[ 4 \alpha ^2 A^{(3)}_{m n^{\prime} n} + \frac{1}{\xi^2(t)}(m^2 A^{(-1)}_{m n^{\prime} n} - B^{(0)}_{m n^{\prime} n} - C^{(1)}_{m n^{\prime} n})
- i \frac{4\alpha}{\xi(t)} \left(  A^{(1)}_{m n^{\prime} n} + B^{(2)}_{m n^{\prime} n} \right)
 \right]~, \label{eq: p0^2_expect}
\\
\langle \Psi_{m^{\prime} n^{\prime}}(t) | H  | \Psi_{mn}(t) \rangle &=&
-\frac{\hbar^2}{2 \mu} \int_0^{2\pi} \int_0^{L(t)} d\rho~\rho~
\Psi^*_{m^{\prime} n^{\prime}}(\rho, \phi, t) \left[ \frac{1}{\rho} \frac{\partial}{\partial \rho} \left( \rho \frac{\partial}{\partial \rho} \right)
+ \frac{\partial^2}{\partial \phi^2} \right] \Psi_{m n}(\rho, \phi, t)
\nonumber
\\
 &=&  \frac{1}{\pi} \frac{ \langle \Psi_{m^{\prime} n^{\prime}}(t) | p^2_{_0}  | \Psi_{mn}(t) \rangle }{2 \mu} ~,
 \label{eq: H_expect}
\end{eqnarray}
where
\begin{eqnarray}
A^{(k)}_{m n^{\prime} n} &=& \int_0^1 ds~s^k~ J_m(x_{mn^{\prime}}s) ~J_m(x_{mn}s)  ~,
\\
B^{(k)}_{m n^{\prime} n} &=& \int_0^1 ds~s^{k}~ J_m(x_{mn^{\prime}}s) ~ \frac{d}{ds}J_m(x_{mn}s) ~,
\\
C^{(k)}_{m n^{\prime} n} &=& \int_0^1 ds~s^{k}~ J_m(x_{mn^{\prime}}s) ~ \frac{d^2}{ds^2}J_m(x_{mn}s) ~,
\end{eqnarray}
are constant coefficients. As one sees, all of the expectation
values are independent of $\phi_0$, the direction we started with.
$A^{(k)}_{m n^{\prime} n}$ is symmetric with respect to two indices
$n^{\prime}$ and $n$, while $B^{(k)}_{m n^{\prime} n}$ and
$C^{(k)}_{m n^{\prime} n}$ do not have any symmetry. It must be
mentioned that $A^{(-1)}_{m n^{\prime} n}$ diverges for $m=0$, but
it is not problematic because the wavefunction is not
$\phi-$dependent for $m=0$ and thus, the term $m^2 A^{(-1)}_{m
n^{\prime} n}$ will not appear in this case.

From eqs. (\ref{eq: A_expect_u}), (\ref{eq: q0_expect}) and (\ref{eq: p0_expect}), one sees that
position and momentum expectation values are zero, irrespective of the shape of the wavefunction,
$\langle \Psi | q_{_0} | \Psi \rangle(t) = 0$ and $\langle \Psi | p_{_0} | \Psi \rangle(t) = 0$;
and from eqs. (\ref{eq: A_expect_u}) and (\ref{eq: H_expect}),
$\langle \Psi | H | \Psi \rangle(t) = \langle \Psi | p^2_{_0} | \Psi \rangle(t) /(2 \pi \mu)$.

\section{Uncertainty Relations}

Let us suppose that one succeed to construct the initial
wavefunction to be the state $\Psi_{mn}({\bf{r}}, 0)$. Such a task
can be done by appropriately superposing instantaneous energy
eigenstates.
%
If the particle is initially in the state $\Psi({\bf{r}}, 0) =
\Psi_{mn}({\bf{r}}, 0)$, then
\begin{eqnarray}
c_{m^{\prime} n^{\prime}} &=& \delta_{m m^{\prime}} \delta_{n n^{\prime}} ~,\label{eq: c_cof_eigen} \\
b_{m^{\prime} n^{\prime}}(t) &=& \delta_{m m^{\prime}} \frac{2}{|
J_{m + 1}(x_{m n}) ~ J_{m + 1}(x_{m n^{\prime}}) |} \exp \left[ -i
x^2_{m n} \frac{1-1/\xi(t)}{4 \alpha} \right] I^*_{m n n^{\prime}
}(t, \alpha)~. \label{eq: b_cof_eigen}
\end{eqnarray}
From eqs. (\ref{eq: geral-sol_1}) and (\ref{eq: c_cof_eigen}), one finds that if $\Psi({\bf{r}}, 0) =
\Psi_{mn}({\bf{r}}, 0)$, then time-evolved wavefunction would be $\Psi_{mn}({\bf{r}}, t)$ in any instant of time.

By using eqs. (\ref{eq: Energy_exp}) and (\ref{eq: b_cof_eigen}), one obtains
\begin{eqnarray} \label{eq: E_mn_expec}
\frac{\langle H \rangle_{_{mn}}(t)}{\langle H \rangle_{_{mn}}(0)}
&=& \left( \frac{a}{L(t)}  \right)^2 \frac{\sum_{n^{\prime}} \left(
\frac{x_{mn^{\prime}}}{J_{m + 1}(x_{m n^{\prime}})} \right)^2 |I_{m
n n^{\prime} }(t, \alpha)|^2 } {\sum_{n^{\prime}} \left(
\frac{x_{mn^{\prime}}}{J_{m + 1}(x_{m n^{\prime}})} \right)^2 |I_{m
n n^{\prime} }(0, \alpha)|^2 } 
\end{eqnarray}
for the energy expectation value. From this relation it is not
obvious how the energy expectation value changes with time, whereas
eq. (\ref{eq: p0^2_expect_diag}) is completely informative in this
connection.

Using eqs.(\ref{eq: q0_expect}), (\ref{eq: p0_expect}), (\ref{eq:
q0^2_expect}) and (\ref{eq: p0^2_expect}),
one obtains
\begin{eqnarray}
\langle q^2_{_0} \rangle_{_{mn}}(t)  &=&
a^2 \xi^2(t) \frac{2 \pi A^{(3)}_{mnn}}{ | J_{m+1}(x_{mn}) |^2 } ~, \label{eq: q0^2_expect_diag}
\\ \nonumber
\\
\langle p^2_{_0} \rangle_{_{mn}}(t) &=& \frac{\hbar^2}{a^2}
\frac{2 \pi}{ | J_{m+1}(x_{mn}) |^2 } \left[ 4 \alpha ^2 A^{(3)}_{mnn}
+ \frac{1}{\xi^2(t)}(m^2 A^{(-1)}_{mnn} - C^{(1)}_{mnn})
\right] \equiv (2\pi \mu) \langle H \rangle_{_{mn}}(t)
~, \label{eq: p0^2_expect_diag}
\\ \nonumber
\\
\Delta q_{_{0, mn}}(t) &\equiv & \sqrt{ \langle q^2_{_0}
\rangle_{_{mn}}(t) - \langle q_{_0} \rangle_{_{mn}} ^2(t) } =
a~\xi(t) \frac{ \sqrt{ 2 \pi A^{(3)}_{mnn} } }{|J_{m+1}(x_{mn})|}
\label{eq: q0_uncer}~,
\\ \nonumber
\\
\Delta p_{_{0, mn}}(t)  &\equiv & \sqrt{ \langle p^2_{_0}
\rangle_{_{mn}}(t) - \langle p_{_0} \rangle_{_{mn}} ^2(t) } =
\frac{\hbar}{a}~\sqrt{ 4\alpha^2  \left( \frac{\Delta q_{_{0,
mn}}(t)}{L(t)} \right)^2 + \frac{1}{\xi^2(t)} \frac{2 \pi (m^2
A^{(-1)}_{mn} - C^{(1)}_{mnn}) }{|J_{m+1}(x_{mn})|^2} } ~,
\label{eq: p0_uncer}
\\ \nonumber
\\
\Delta q_{_{0, mn}}(t) \Delta p_{_{0, mn}}(t) &=& \frac{\hbar}{2} ~
\frac{4\pi}{|J_{m+1}(x_{mn})|^2} \sqrt{ (m^2 A^{(-1)}_{mnn} - C^{(1)}_{mnn})
A^{(3)}_{mnn} + ( 2\alpha  A^{(3)}_{mnn} )^2~\xi^2(t) }
\label{eq: uncer_prod}
\end{eqnarray}
where, for brevity the expectation values have been shown as
$\langle ... \rangle_{_{mn}}(t)$ instead of $\langle
\Psi_{mn}(t)|...|\Psi_{mn}(t) \rangle$; and $B^{(0)}_{mnn} = 0$. In
appendix, the solutions of $A^{(k)}_{mnn}$ and $C^{(1)}_{mnn}$ will
be given in terms of the hypergeometric and the regularized
hypergeometric functions.

Considering the above equation, one finds
\begin{enumerate}[i)]

\item $\Delta q_{_{0, mn}}(t)$ changes linearly with time, it increases for expansion while decreases for contraction. This is an expected result.

\item There is an additional time- and $\hbar$-independent contribution to $\Delta p_{_{0, mn}}$ in comparison
with the stationary boundary. The second term under the radical sign is time-dependent and
decreases with expansion while increases in contraction. For an
expanding circle, $\Delta p_{_{0, mn}}(t)$ becomes constant in the
limit $t \rightarrow \infty$.

\item Eq. (\ref{eq: uncer_prod}) shows that the product of the
uncertainties is time-dependent, a result which is absent in
stationary systems. According to the general uncertainty principle
$\Delta q_i \Delta p_i \geq \hbar/2$, where $q_i$ and $p_i$ are the
non-commuting $i-$component of the conjugate variables $\bf{r}$ and
$\bf{p}$. This means that the quantity $4\pi \sqrt{ (m^2
A^{(-1)}_{mnn} - C^{(1)}_{mnn}) A^{(3)}_{mnn}}/|J_{m+1}(x_{mn})|^2$,
which is the product of uncertainties in the corresponding
stationary system, must be greater than $1$.

\end{enumerate}

\section{Discussion}
\begin{figure}
\centering
\includegraphics[width=12cm,angle=-90]{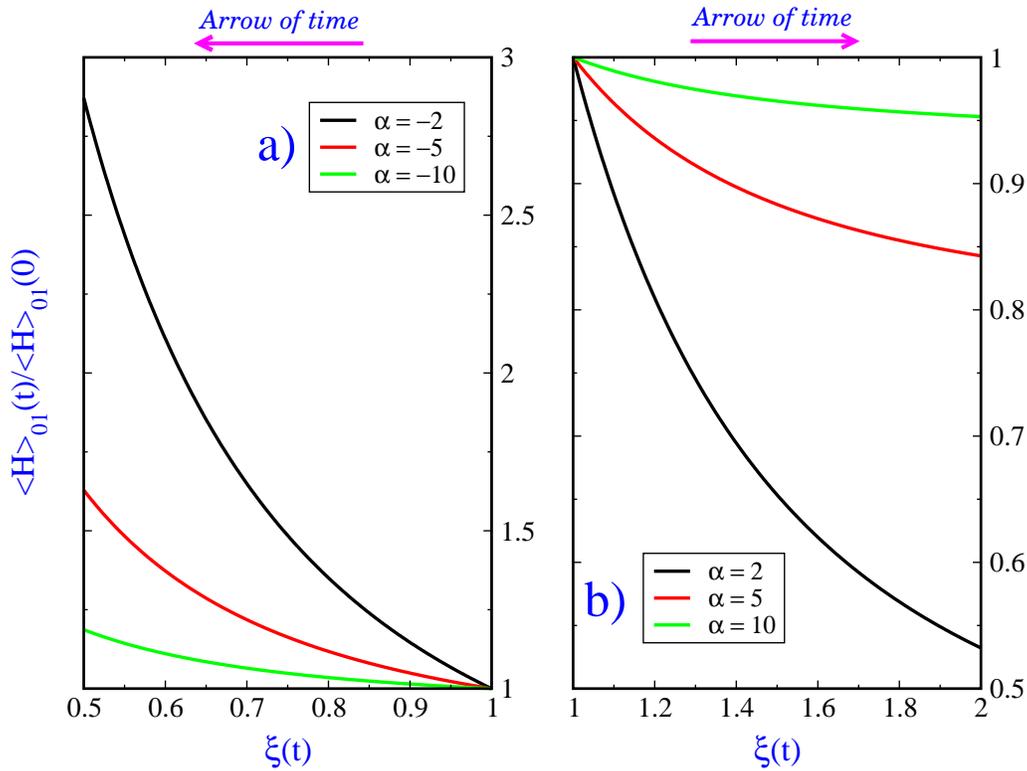}
\caption{(Color online) $\langle H \rangle_{_{01}}(t)/\langle H \rangle_{_{01}}(0)$ versus $\xi(t)$,
for (a) contracting and b) expanding circular boxes. Expectation value of the energy of the confined particle
increases (decreases) for a contracting (an expanding) box.}
\vspace*{0.5cm}
\label{fig: E_expectation}
\end{figure}

As eq.(\ref{eq: p0^2_expect_diag}) shows and fig. \ref{fig: E_expectation} confirms, expectation value of the energy
decreases with time for an expanding box while increases for a contracting one, as time elapses.
This is an understandable result in the context
of the old quantum theory \cite{Pi-AJP-1989}, as is reasonable by considering
the uncertainty relations (\ref{eq: q0_uncer}) and (\ref{eq: p0_uncer}) \cite{Wi-JPA-1983}.


\begin{figure}
\centering
\includegraphics[width=12cm,angle=-90]{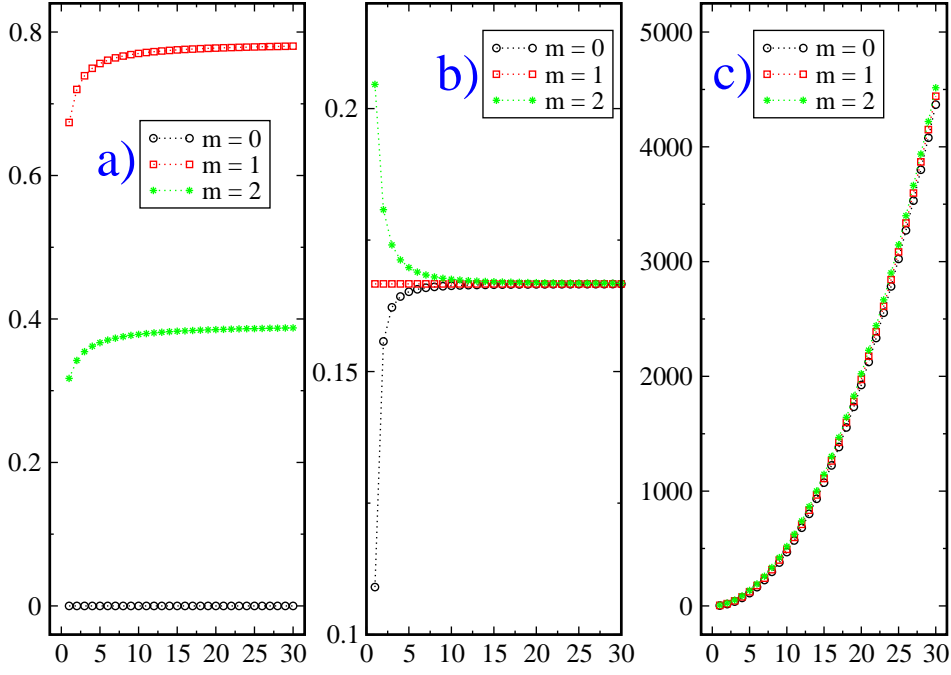}
\caption{(Color online) (a) $A^{(-1)}_{mnn}/|J_{m+1}(x_{mn})|^2$, (b) $A^{(3)}_{mnn}/|J_{m+1}(x_{mn})|^2$ and (c) $|C^{(1)}_{mnn}|/|J_{m+1}(x_{mn})|^2$
versus quantum number $n$ for different values of quantum number $m$.}
\vspace*{0.5cm}
\label{fig: AC_coefs}
\end{figure}

Fig. \ref{fig: AC_coefs} shows that in the presented region,
$A^{(-1)}_{mnn}/|J_{m+1}(x_{mn})|^2$ and
$A^{(3)}_{mnn}/|J_{m+1}(x_{mn})|^2$ becomes constant for $n \geq
10$, while $|C^{(1)}_{mnn}|/|J_{m+1}(x_{mn})|^2$ increases with $n$.
In addition, $A^{(3)}_{mnn}/|J_{m+1}(x_{mn})|^2$ is approximately
independent of quantum number $m$ for $n \geq 10$, while
$C^{(1)}_{mnn}/|J_{m+1}(x_{mn})|^2$ is approximately the same for
all values of $n$ and dominates for $n \geq 3$. From these
considerations; and eqs. (\ref{eq: q0^2_expect_diag}) and (\ref{eq:
p0^2_expect_diag}), one finds that at a given time and a given wall
velocity, $\langle q^2_{_0} \rangle_{_{mn}}$ becomes independent of
$n$ and $m$ for $n \geq 10$, while $\langle p^2_{_0}
\rangle_{_{mn}}$ is independent of $m$, but increases with $n$.

For numerical calculations of the wavefunction, we define two new quantities $\lambda_{m
n} = 2\pi a/x_{m n}$ and $\nu_{m n} = E_{m n}/(2\pi\hbar)$ and from
them new dimensionless position and time coordinates $\eta =
\rho/\lambda_{m n}$ and $T = \nu_{m n} t$; and dimensionless radial
probability density $\varrho_{m n}(\eta, T) = \lambda_{m n}^2 \eta
|R(\eta, T)|^2$, where $R(\eta, T)$ stands for the radial part of
the wavefunction. With these quantities, the location of the wall is
determined from the relation
\begin{eqnarray*}
\xi(T) &=& 1 + 2\pi \frac{\alpha}{\alpha^2_{mn}}T~,
\end{eqnarray*}
where,
\begin{eqnarray*}
\alpha_{mn} &=& \frac{\mu a}{2\hbar} v_{mn} = \frac{\mu a}{2\hbar}
\frac{\hbar x_{mn}}{\mu a} = \frac{x_{mn}}{2}~,
\end{eqnarray*}
corresponds to the wall velocity that is equal to the initial
classical velocity of the confined particle.

\begin{figure}
\centering
\includegraphics[width=12cm,angle=-90]{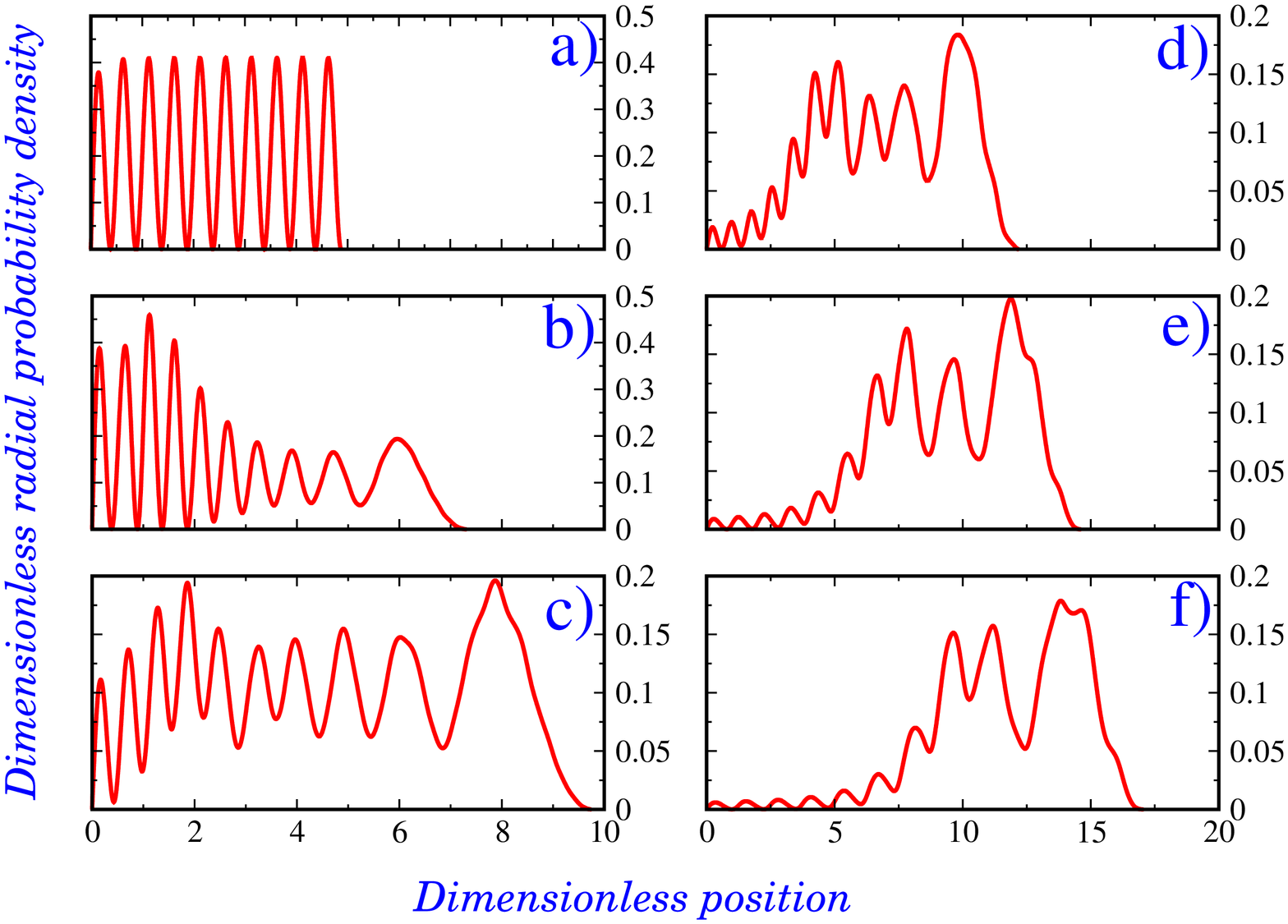}
\caption{(Color online) Dimensionless radial probability density
$\varrho_{0, 10}(\eta, T)$ for $\alpha = \alpha_{0, 10}$ versus
dimensionless position coordinate $\eta$ at the instant of time when
the wall arrives at (a) $\xi = 1$, (b) $\xi = 1.5$, (c) $\xi = 2$,
(d) $\xi = 2.5$, (e) $\xi = 3$ and (f) $\xi = 3.5$.}
\vspace*{0.5cm}
\label{fig: wave_r_10}
\end{figure}

We have plotted $\varrho_{m n}(\eta, T)$  in fig. \ref{fig: wave_r_10}
for a particle initially in the state $u_{0,10}$, versus
$\eta$ at different times.

\begin{figure}
\centering
\includegraphics[width=12cm,angle=-90]{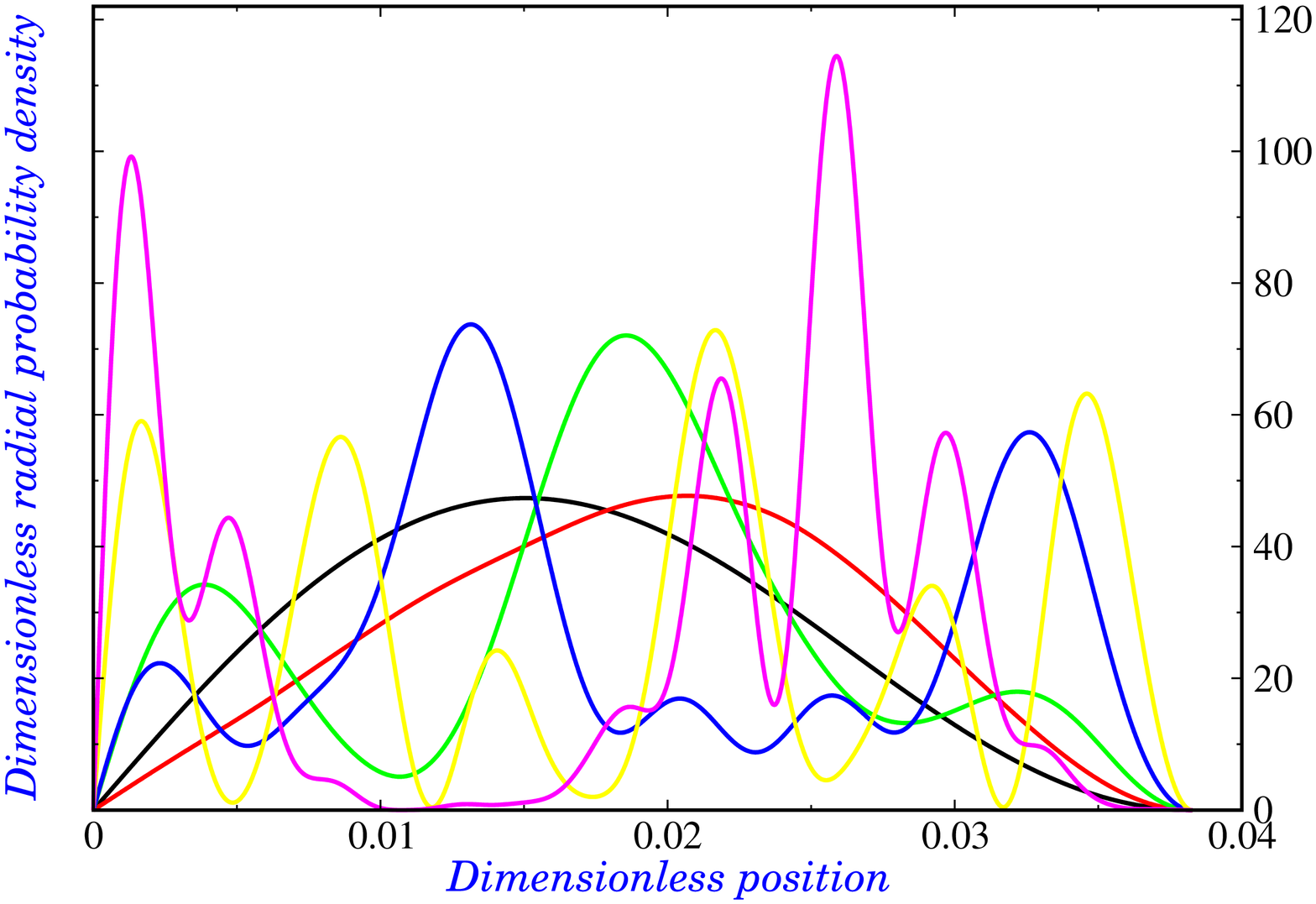}
\caption{(Color online) Dimensionless radial probability
density $\varrho_{0, 1}(\eta, T)$ versus dimensionless position
coordinate $\eta$ at the instant of time when the wall arrives at
$\xi = 0.1$, for six different values of contraction rate; $\alpha =
-0.01 ~ \alpha_{0, 1}$ (black curve), $\alpha = - \alpha_{0, 1}$ (red
curve), $\alpha = -5 ~ \alpha_{0, 1}$ (green curve), $\alpha = -10
 ~ \alpha_{0, 1}$ (blue curve), $\alpha = -15 ~ \alpha_{0, 1}$ (yellow
curve) and $\alpha = -20 ~ \alpha_{0, 1}$ (magenta curve).
For a rapidly moving boundary, the sudden approximation does not work for
a contracting box, but, in the opposite limit the process is adiabatic.}
\vspace*{0.5cm}
\label{fig: wave_r_1}
\end{figure}

\begin{figure}
\centering
\includegraphics[width=12cm,angle=-90]{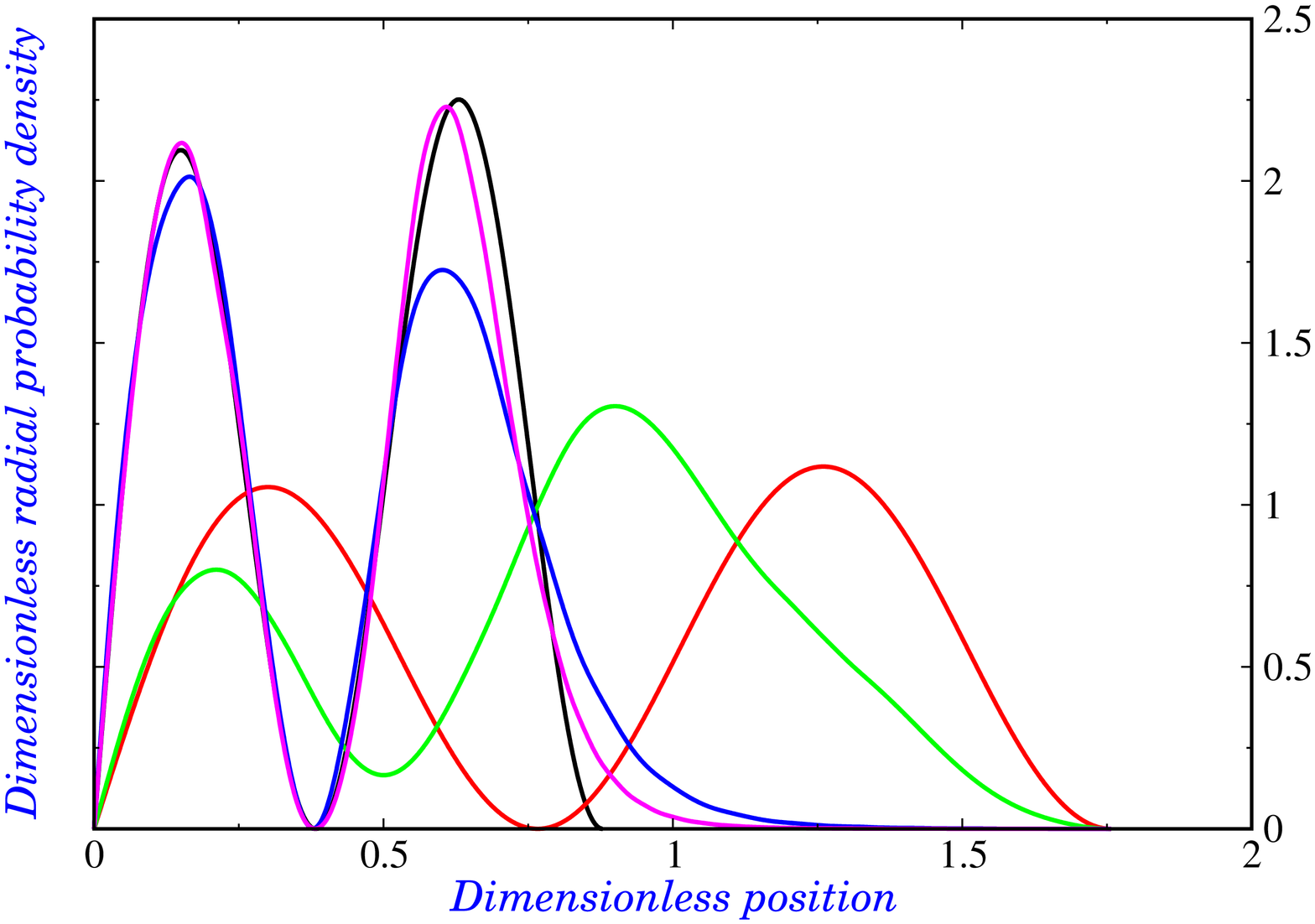}
\caption{(Color online) Dimensionless radial probability
density $\varrho_{0, 2}(\eta, T)$ versus dimensionless position
coordinate $\eta$ at the instant of time when the wall arrives at
$\xi = 2$, for five different values of expansion rate; $\alpha = 0$
(black curve), $\alpha = 0.01 ~ \alpha_{0, 2}$ (red curve), $\alpha =
\alpha_{0, 2}$ (green curve), $\alpha = 5 ~ \alpha_{0, 2}$ (blue
curve) and $\alpha = 10 ~ \alpha_{0, 2}$ (magenta curve).
Both the adiabatic and the sudden approximation work for an expanding box.}
\vspace*{0.5cm}
\label{fig: wave_r_2}
\end{figure}

A system undergoes an adiabatic evolution when $t_{\text{i}} \ll
t_{\text{e}}$, where $t_{\text{i}}$ ($t_{\text{e}}$) is the
time-scale over which internal (external) variables of the system
changes \cite{Gr-book-1994}. For our particle-in-a-box,
$t_{\text{e}} = a/u$ and $t_{\text{i}} = a/v_{m n}$. Figures
\ref{fig: wave_r_1} and \ref{fig: wave_r_2} show that the particle,
initially in an instantaneous energy eigenstate will remain in the
same eigenstate when $|\alpha| \ll \alpha_{m n}$, as a consequence
of the adiabatic approximation. One sees that the sudden
approximation does not work for a contracting well
\cite{Pi-AJP-1989}.



\begin{figure}
\centering
\includegraphics[width=12cm,angle=-90]{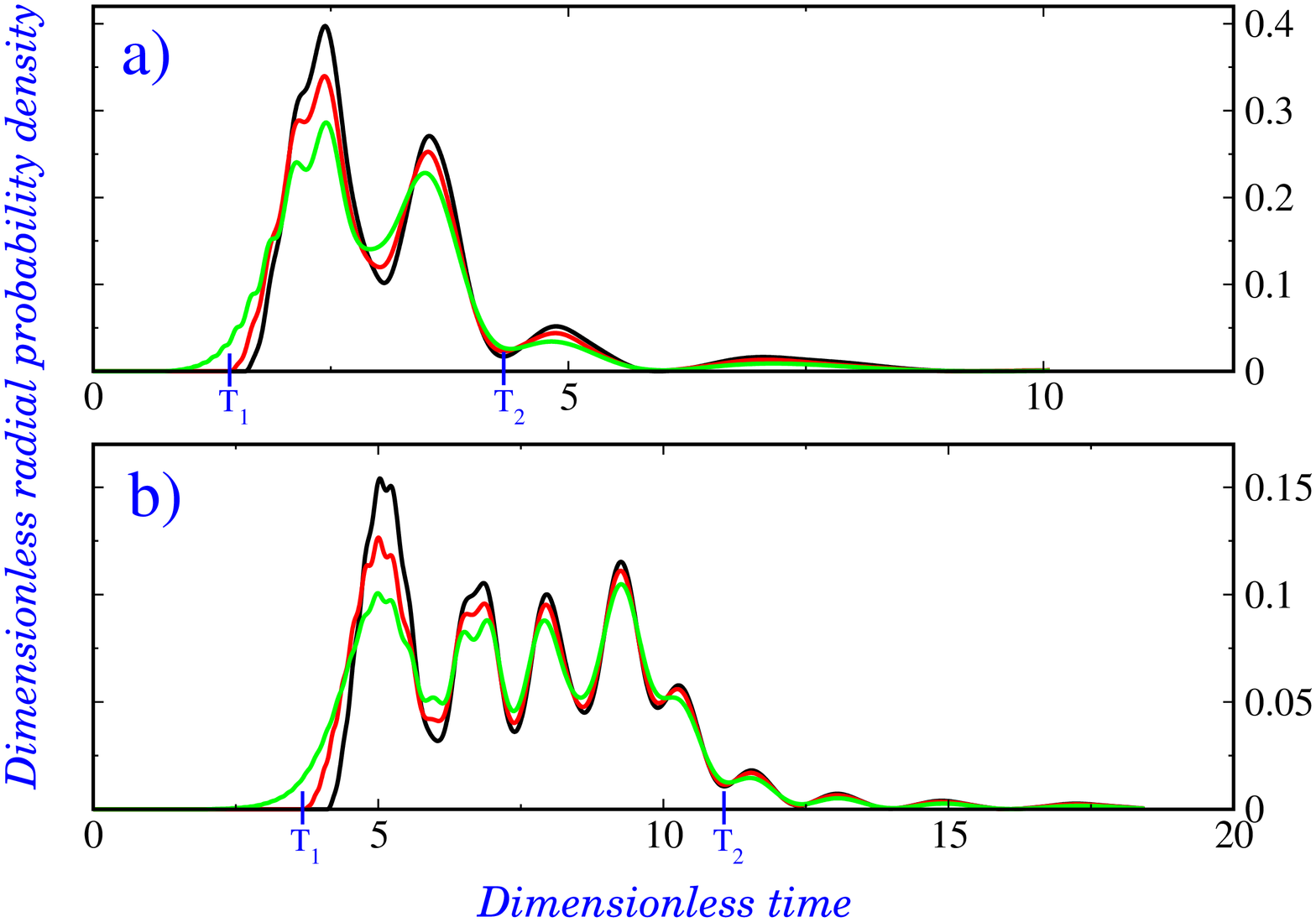}
\caption{(Color online) Dimensionless radial probability
density $\varrho_{m n}(\eta, T)$ for a particle initially in the state
(a) $u_{0,6}$ and (b) $u_{0,15}$, versus dimensionless time
coordinate $T$ at dimensionless observation point $\eta^{(0)} =
x_{mn}/\pi$, for three different values of velocity parameter;
$\alpha = 0.9 ~ \alpha_{m n}$ (black curve), $\alpha = \alpha_{m n}$
(red curve), $\alpha = 2 ~ \alpha_{m n}$ (green curve). $T_1$ and
$T_2$ are dimensionless classical flight times from the front and
back edges of the circle to the dimensionless observation point.
As the velocity of the moving boundary increases, the fringe visibility
reduces.}
\vspace*{0.5cm}
\label{fig: wave_t}
\end{figure}

Figure \ref{fig: wave_t} represents density probability
versus time in a given observation point.

$T_1$ and $T_2$ are dimensionless classical flight times from the
front and back edges of the circle to the dimensionless observation
point. In contrast to the classical mechanics, when $\alpha >
\alpha_{mn}$, one sees a non-monotonous increasing behavior of the
density for $T < T_1$. A non-monotonous decreasing behavior is seen
for $T > T_2$, irrespective of the wall velocity. Difference between
the height adjacent extremums (visibility) decreases with $\alpha$.
The constructive interference with the reflected components from the
moving boundary for $\alpha < \alpha_{mn}$, increases the fringe
visibility.

At long times, the behavior of the density in the observation point
is approximately the same for all values of the velocity parameter
$\alpha$.

This can be understood as follows.
From eqs. (\ref{eq: instan-waves}), (\ref{eq: norm_waves}), (\ref{eq: geral-sol_1}) and (\ref{eq: integral}) one obtains
\begin{eqnarray}
|R(\rho_{_0}, t)| &=& \frac{2}{|J_{m+1}(x_{mn})|} \frac{\sqrt{2}}{L(t)} \sum_{n^{\prime}}
\frac{|I_{mnn^{\prime}}(0, \alpha)|}{(J_{m+1}(x_{mn^{\prime}}))^2} J_m \left(x_{mn^{\prime}} \frac{\rho_{_0}}{L(t)} \right)
\end{eqnarray}
for the radial part of the wavefunction at an arbitrary observation point $\rho = \rho_{_0}$. It is approximately,
\begin{eqnarray}
|R(\rho_{_0}, t)| &\simeq & \frac{2}{|J_{m+1}(x_{mn})|} \frac{\sqrt{2}}{ut} \sum_{n^{\prime}}
\frac{|I_{mnn^{\prime}}(0, \alpha)|}{(J_{m+1}(x_{mn^{\prime}}))^2} J_m \left(x_{mn^{\prime}} \frac{\rho_{_0}}{ut} \right)
\end{eqnarray}
at long times and ultimately vanishes. So, the difference between
the values of $|R(\rho_{_0}, t)|$ for two different values of
velocity $u$ becomes vanishingly small at this limit


\section{Summary}

Exact solutions of the Schr\"{o}dinger equation for a particle in a
circular impenetrable box with a moving wall in uniform motion,
contain a coordinate-dependent phase $\exp\left[ i\frac{\mu
u}{2\hbar} \frac{\rho^2}{L(t)} \right]$. This phase factor does not
appear in the corresponding stationary boundary problem and its
significance has been already emphasized \cite{{Ma-JPA-1992},
GrDoMo-Phase} in 1D systems. Propagator of the problem was
constructed and the matrix elements of the position, the momentum
and the energy observables were derived with respect to the exact
solutions. It was seen that the uncertainty product increases with
time due to a time- and velocity-dependent term; and expectation
value of the energy increases (decreases) with time, for a
contracting (an expanding) boxes which is consistent with the
uncertainty relations. Transients corresponding to the operation of
expansion or contraction which is quite common in cold atom traps,
were studied carefully. The density profile in the time, in a given
location, resembles the diffraction-in-time pattern observed in a
suddenly released particle, but it actually shows an enhancement of
the visibility of the fringes, the 2D version of the EDIT (Enhanced
diffraction in time) reported in 1D in \cite{CaMuCl-PRA-2008}. We
have provided a full characterization of the quantum transients in
an expanding/contracting cylindrical box resulting from the
breakdown of adiabaticity.
We close by noting that a shortcut to the adiabaticity
\cite{ChRuScCaGuMu-PRL-2010} can be implemented to suppress quantum
transients in this system \cite{CaBo-arxiv-2012}.
\\
\\

{\bf{Acknowledgment}} We are grateful to A. del Campo for valuable
suggestions; and to S. Fallahi and A. Azarm for corrections. The
author also thanks the anonymous referees for their comments and
suggestions. Financial support from the University of Qom is
acknowledged.


\section{Appendix}

Here, we write integrals $A^{(-1)}_{mnn}$, $A^{(3)}_{mnn}$ and $C^{(1)}_{mnn}$ in terms of generalized and regularized generalized hypergeometric functions:
\begin{eqnarray} \label{eq: A(-1)}
A^{(-1)}_{mnn} &=& \frac{1}{2m} \int_0^1 ds~ J_m(x_{mn}s) ( J_{m+1}(x_{mn}s) + J_{m-1}(x_{mn}s) )
\nonumber
\\
&=&
4^{-m} (2m-1)! ~{_2{\bar{F}}_3} \left( m, m+\frac{1}{2};  m+1, m+1, 2m+1; -(x_{mn})^2  \right) ~~~~~~ \text{if } m \neq 0~,
\end{eqnarray}
%

\begin{eqnarray}
A^{(3)}_{mnn} &=&
\begin{cases}
  4^{-m} m(m+1) (x_{mn})^2 (2m-1)! ~{_2{\bar{F}}_3} \left( m+\frac{1}{2}, m+2; m+1, m+3, 2m+1; -(x_{mn})^2  \right)
& \text{if } m \neq 0,
  \\
  \frac{1}{4} ~{_2F_3} \left( \frac{1}{2}, 2; 1, 1, 3; -(x_{0n})^2 \right) & \text{if } m=0,
\end{cases}
\end{eqnarray}
\begin{eqnarray}
C^{(1)}_{mnn} &=& - \int_0^1 ds~ s \left( \frac{dJ_m(x_{mn}s)}{ds}  \right)^2
\nonumber
\\
&=& -
\frac{(x_{mn})^2}{4} \bigg[ -\frac{ (x_{mn})^{2m}~{_3F_4}
\left( m+\frac{1}{2}, m+1, m+1; m, m+2, m+2, 2m+1; -(x_{mn})^2
\right) }{ 4^m (m-1)!~ m!}
\nonumber
\\
&+& \frac{ 2 J_{m-2}(x_{mn}) J_{m-1}(x_{mn}) \left( 4m(m^2-1)+(x_{mn})^2(1-2m) \right) }{ (x_{mn})^3 }
+ (J_{m-2}(x_{mn}))^2 \frac{(x_{mn})^2- 2m(m+1)}{(x_{mn})^2}
\nonumber
\\
&+& (J_{m-1}(x_{mn}))^2 \left( \frac{1}{2} - \frac{4m(m-1)(2m^2-2-(x_{mn})^2)}{(x_{mn})^4} \right)
+ \frac{ (J_{m+1}(x_{mn}))^2 }{2}
\bigg]~~~~\text{if } m \neq 0
\end{eqnarray}
and,
\begin{eqnarray}
C^{(1)}_{mnn} &=& - \frac{1}{2} x_{0n} (J_1(x_{0n}))^2 ~~~~ \text{if } m = 0 ~,
\end{eqnarray}
where the generalized hypergeometric function ${_pF_q}(a_1, ...,
a_p; b_1, ..., b_q; z)$ in terms of the Pochhammer symbol
\cite{Arfken-book-2005},
\begin{eqnarray*}
(a)_n &=& \frac{(a+n-1)!}{(a-1)!}~, ~~~ (a)_0 = 1,
\end{eqnarray*}
becomes
\begin{eqnarray*}
{_pF_q}(a_1, ..., a_p; b_1, ..., b_q; z) &=& \sum_{k=0}^{\infty}
\frac{(a_1)_k~...~(a_p)_k}{(b_1)_k~...(b_q)_k} \frac{z^k}{k!}~,
\end{eqnarray*}
and the regularized generalized hypergeometric function ${_p{\bar{F}}_q}(a_1, ..., a_p; b_1, ..., b_q; z)$ is
\begin{eqnarray*}
{_p{\bar{F}}_q}(a_1, ..., a_p; b_1, ..., b_q; z) &=& \frac{ {_pF_q}(a_1, ..., a_p; b_1, ..., b_q; z) } { \Gamma(b_1) ... \Gamma(b_q) } ~,
\end{eqnarray*}
and $\Gamma(z) = \int_0^{\infty} t^{z-1}e^{-t} dt$ is the Gamma function.
Recurrence relations of the Bessel functions have been used in (\ref{eq: A(-1)}) to re-write the integrand, which is more appropriate for numerical calculations.


\end{document}